\documentclass[onecolumn,11pt]{article}
\usepackage[top=2.54cm, bottom=2.54cm, left=2.54cm, right=2.54cm,a4paper]{geometry}%\documentclass[prl,letterpaper,twocolumn]{revtex4}

\usepackage{amsmath,amsthm}
\usepackage{bm}  % Define \bm{} to use bold math fonts
\usepackage{bbm}
\usepackage{verbatim}
\usepackage{xcolor}

\usepackage{qcircuit}

\usepackage[backend=bibtex,sorting=none,style=phys,biblabel=brackets,backref=false,bibencoding=utf8,maxnames=20,eprint=true]{biblatex}
\makeatletter
\pretocmd{\blx@head@bibintoc}{\phantomsection}{}{\ddt}
\makeatother
\DefineBibliographyStrings{english}{%
  backrefpage = {page},% originally "cited on page"
  backrefpages = {pages},% originally "cited on pages"
}
\usepackage{titlesec}

\titleformat*{\section}{\bfseries}
\titleformat*{\subsection}{\normalsize\bfseries}
\titleformat*{\subsubsection}{\bfseries}
\titleformat*{\paragraph}{\large\bfseries}
\titleformat*{\subparagraph}{\large\bfseries}

\titlespacing\section{0pt}{12pt plus 4pt minus 2pt}{2pt plus 2pt minus 2pt}

\usepackage[bookmarks,colorlinks,breaklinks]{hyperref}  % PDF hyperlinks, with coloured links
\definecolor{dullmagenta}{rgb}{0.4,0,0.4}   % #660066
\definecolor{darkblue}{rgb}{0,0,0.4}
\hypersetup{linkcolor=red,citecolor=blue,filecolor=dullmagenta,urlcolor=darkblue} % coloured links

\usepackage[bitstream-charter,cal=cmcal]{mathdesign}
\usepackage[T1]{fontenc}

\usepackage{amsbsy,verbatim,graphicx}

\newcommand{\ket}[1]{|#1\rangle}

\newcommand{\bra}[1]{\langle #1|}

\newcommand{\ketbra}[1]{\ket{#1}\bra{#1}}

\newcommand{\id}{\mathbbm 1}

\def\cM{\mathcal M}
\def\cN{\mathcal N}

\def\X{{\mathtt{X}}}
\def\Z{{\mathtt{Z}}}
\def\I{{\mathtt{I}}}
\def\Y{{\mathtt{Y}}}
\newcommand{\ubar}[1]{\text{\b{$#1$}}}
\def\uz{\ubar{0}}
\def\uo{\ubar{1}}

\usepackage{xfrac}

\usepackage{setspace}
\usepackage{titling}

\posttitle{\par\end{center}}
\predate{}
\postdate{}
\begin{document}
%\author{Joseph M.~Renes}
\author{
{\normalsize 
\'Alvaro Piedrafita\textsuperscript{1,2}\thanks{AP completed most of this work while at ETH Z\"urich.} and 
Joseph M.\ Renes\textsuperscript{1}}\\
\emph{\normalsize  
\textsuperscript{1}Institute for Theoretical Physics, ETH Z\"urich, Switzerland}\\
\emph{\normalsize \textsuperscript{2}Qusoft and CWI, Amsterdam, The Netherlands}
}

%\author{{\normalsize \'Alvaro Piedrafita and Joseph M.~Renes}\\
%{ \normalsize \emph{Qusoft and CWI, Amsterdam, The Netherlands\\
%Institute for Theoretical Physics, ETH Z\"urich, 8093 Z\"urich, Switzerland}}
%\thanks{Institute for Theoretical Physics, ETH Zurich, 8093 Z\"urich, Switzerland}
%}

%\affiliation{Institut f\"ur Angewandte Physik, Technische Universit\"at Darmstadt, Hochschulstr.~4a, 64289 Darmstadt, Germany}
%\affiliation{Institut f\"ur Theoretische Physik, ETH Zurich, Wolfgang-Pauli-Str.\ 27, 8093 Z\"urich, Switzerland}

\title{\large {\bf Reliable channel-adapted error correction: Bacon-Shor code recovery from amplitude damping}}
%Approximate Quantum Error Correction via Complementary Observables}

\date{\vspace{-\baselineskip}}
%\vspace{-1.3cm}

%\maketitle

\maketitle

\begin{abstract}
We construct two simple error correction schemes adapted to amplitude damping noise for Bacon-Shor codes and investigate their prospects for fault-tolerant implementation. 
Both consist solely of Clifford gates and require far fewer qubits, relative to the standard method, to achieve correction to a desired order in the damping rate. 
The first, employing one-bit teleportation and single-qubit measurements, needs only one fourth as many physical qubits, while the second, using just stabilizer measurements and Pauli corrections, needs only half.  
We show that existing fault-tolerance methods can be employed for the latter, while the former can be made to avoid potential catastrophic errors and can easily cope with damping faults in ancilla qubits. 
\end{abstract}
\vspace{0.5\baselineskip}

%The first, based on one-bit teleportation and single-qubit measurements, needs only one fourth as many, while the second needs only half as many,  correction. 
%The second needs half as many qubits, but is also a syndrome-based scheme in that the only required quantum operations are stabilizer measurements and Pauli corrections.

\section{Introduction}
The performance of quantum error correction over a given channel can be substantially improved by adapting the decoding algorithm to the channel. 
For the amplitude damping channel, which describes the effects of energy relaxation, Leung \emph{et al.}\ illustrate this quite dramatically with a four qubit code capable of correcting single damping errors~\cite{leung_approximate_1997}, one qubit fewer than the minimum number needed to recover from an arbitrary single-qubit error~\cite{bennett_mixed-state_1996,knill_theory_1997}. 
However, there is no guarantee that the performance available to a given code and channel can be realized by simple correction circuits.
For instance, although the Shor code~\cite{shor_scheme_1995} can in principle correct two amplitude damping errors~\cite{gottesman_stabilizer_1997}, standard recovery based on stabilizer measurement and Pauli recovery operations will correct only one. 
And the problem of finding simple high-performance decoders is compounded when requiring the decoder to tolerate noise during its implementation. 
Nonetheless, reducing the qubit overhead needed for error correction would be tremendously useful in near term efforts to construct quantum information processors.

In this paper we investigate the performance of simple decoding schemes for the Bacon-Shor code family~\cite{bacon_operator_2006} adapted to amplitude damping noise. 
The $(n,m)$ Bacon-Shor code combines two classical repetition codes, a length-$n$ code for phase flips and a length-$m$ code for bit flips, and by standard stabilizer recovery can correct up to $\min(\lfloor(n-1)/2\rfloor,\lfloor(m-1)/2\rfloor)$ arbitrary single-qubit errors. 
Duan \emph{et al.}\ observe that for amplitude damping noise the $(t+1,t+1)$ code exactly satisfies the error-correcting conditions to order $t$ in the damping probability $\gamma$~\cite{duan_multi-error-correcting_2010}, meaning Bacon-Shor codes can in principle correct twice as many damping errors as arbitrary single-qubit errors. 
Here we present an error correcting procedure using only Clifford operations that achieves this performance, as well as a syndrome-based procedure (requiring only stabilizer measurements) which perfectly recovers $(t+1,2t+1)$ codewords to order $t$.

Furthermore, we also investigate the prospects for fault-tolerant implementation of the two correction schemes. 
Our syndrome correction method also corrects the Pauli twirl~\cite{bennett_mixed-state_1996} of the amplitude damping channel, which makes it amenable to the use of standard fault-tolerant gadgets. 
The Clifford correction does not, however, complicating the analysis. 
Nevertheless, we find that while single, ill-timed damping events can be catastrophic for direct implementation of the correction procedure, this problem can be avoided by suitable alterations. 
Additionally, we show that damping errors in ancillas used for error correction propagate into the data as phase errors, and are therefore correctable by subsequent correction cycles.  
This would seem to require the use of the $(2t+1,2t+1)$ code for syndrome correction, wiping out all qubit savings relative to standard decoding. 
But the $(2t+1,t+1)$ code would suffice for Clifford correction, meaning it could reliably recover from combined amplitude damping and dephasing noise (often characterized as $T_1$ and $T_2$ decay times, respectively~\cite{nielsen_quantum_2000}) using half as many qubits as the standard method. 

Channel adapted error correction has been studied by several authors. 
Fletcher \emph{et al.}\ consider the problem of finding structured decoders in~\cite{fletcher_structured_2008} and \cite{fletcher_channel-adapted_2008}.
The latter, which is specifically focused on amplitude damping, extends the construction of Leung \emph{et al.}\ to higher rates and gives a stabilizer-based recovery, but does not consider fault-tolerance. 
They do describe a stabilizer-based decoder for the Shor code which outperforms standard decoding, but it does not correct all errors to second order. 
Several further codes adapted to the amplitude damping channel have been found in~\cite{duan_multi-error-correcting_2010,shor_high_2011,grassl_quantum_2014,jackson_concatenated_2016}. 
Indeed, Duan \emph{et al.}\ observe that the $(t+1,t+1)$ code can correct damping to order $t$, but do not give an explicit error correction scheme. 
A comparison of various short codes subject to generalized amplitude damping is given in \cite{cafaro_approximate_2014}.  
Meanwhile, fault-tolerance of Bacon-Shor codes has been analyzed for different noise models. 
Ralph \emph{et al.}\ considered protection of optically-encoded quantum information against photon loss (erasure)~\cite{ralph_loss-tolerant_2005}, and this approach was recently analyzed for constructing a quantum repeater~\cite{muralidharan_ultrafast_2014}. 
The effectiveness of concatenated and plain Bacon-Shor codes against Pauli noise is considered in~\cite{aliferis_subsystem_2007,cross_comparative_2009} and \cite{napp_optimal_2013,brooks_fault-tolerant_2013}, respectively. 

We have structured the presentation of our results as follows. 
The next section details the amplitude damping channel and Bacon-Shor codes. 
We describe the two correction methods in \S\ref{sec:ec} and prospects for fault-tolerant implementation in \S\ref{sec:ft}.

\section{Setup}
\subsection{Amplitude damping channel}
The amplitude damping channel describes the process of energy relaxation of a system from an excited state to its ground state.  
Supposing that the probability for decay is $\gamma$ and we are only interested in the two-dimensional space of the excited and ground states, the action of the channel $\cN_\gamma$ is given by the following two Kraus operators
\begin{align}
\label{eq:krausops}
A_0&=\begin{pmatrix}1 & 0\\ 0& \sqrt{1-\gamma}\end{pmatrix}\qquad \text{and}\qquad
A_1=\begin{pmatrix}0 & \sqrt{\gamma}\\ 0& 0\end{pmatrix}.
\end{align}
An arbitrary state $\rho$ is transformed into $\cN_\gamma(\rho)=A_0\rho A_0^\dagger+A_1\rho A_1^\dagger$. 

Unlike a Pauli channel, neither of the Kraus operators has trivial effect on the input state. 
Damping, the action of $A_1$, is of course linear in the probability $\gamma$ (acting on the density operator, not state vector). 
Meanwhile, the nondamping operator can be expressed as a superposition of identity $\mathtt{I}$ and phase flip $\Z=\sigma_z$,
\begin{subequations}
\label{eq:A0}
\begin{align}
A_0&=\tfrac12\left((1+\sqrt{1-\gamma})\mathtt{I}+(1-\sqrt{1-\gamma})\Z\right)\\
&=(1-\tfrac\gamma4)\mathtt{I}+\tfrac\gamma4\Z+O(\gamma^2).
\end{align}
\end{subequations}
Thus, by the usual discretization argument~\cite{shor_scheme_1995,gottesman_introduction_2010}, phase error correction is sufficient to reverse the action of $A_0$ to second order in $\gamma$.  

It will also be useful to consider the Pauli twirl of $\cN_\gamma$~\cite{bennett_mixed-state_1996}, the Pauli channel resulting from randomizing the orientation of the input and output in the following manner: 
\begin{align}
\cN^{\text{Pauli}}_\gamma(\rho)=\tfrac14 \sum_{k=0}^{3} \sigma_k^\dagger \cN_\gamma(\sigma_k\rho\sigma_k^\dagger)\sigma_k\,,
\end{align}
where we have used the usual notation for the Pauli operators for convenience, including $\sigma_0=\I$. 
Amplitude damping is of course already covariant with respect to rotations about the $z$ axis, so only the $\X=\sigma_x$ randomization has any effect here. 
It is easy to work out that the probabilities of the various Pauli operators in $\cN^{\text{Pauli}}_\gamma$ are just $\tfrac14\gamma$ for both $\X$ and $\Y$ and $\tfrac14(2-\gamma\pm 2\sqrt{1-\gamma})$ for $\I$ and $\Z$, respectively.  
Expanding the latter to second order, we have $1-\tfrac12\gamma-\tfrac1{16}\gamma^2$ for $\I$ and $\tfrac1{16}\gamma^2$ for $\Z$, corresponding to the contribution of phase errors by $A_0$ at second order. 

\subsection{Bacon-Shor codes}
\label{sec:BS}
Bacon-Shor codes are stabilizer-based subsystem codes built from two classical repetition codes~\cite{bacon_operator_2006}.
The $(n,m)$ code encodes a single qubit into $nm$ physical qubits, which can be thought of as arranged on a rectangular $n\times m$ lattice. 
The stabilizer generators of the code are then given by the products of $\X$ operators on all qubits in any two neighboring rows and the products of $\Z$ operators on all qubits in two neigboring columns.
Additional operators are needed to fix the ``gauge'' of the code.
For our purposes we will be interested in the $\Z$ gauge, where the gauge operators are $\Z$ operators on pairs of qubits anywhere in a row. 
Note that the $\Z$-type stabilizers are products of these $\Z\Z$ gauge operators. 
In the complementary $\X$ gauge, the additional stabilizers are $\X$ operators on pairs of neighboring qubits anywhere in a column; these similarly subsume the $\X$-type stabilizers. 
Our convention is that the logical operators $\bar \X$ and $\bar\Z$ of the code are just the row of $\X$ operators along the top and the column of $\Z$ operators along the left, respectively. 
Let us denote these operators by $\X^{\text T}$ and $\Z^{\text L}$, respectively. 

Unsurprisingly, the Shor code is indeed the $(3,3)$ $\Z$-gauge code (with logical $\X$ and $\Z$ swapped) and it happens that the Leung \emph{et al.}\ code is the $(2,2)$ $\Z$-gauge code. 
By using the stabilizers to separately perform standard correction of the repetition code for bit flips and phase flips, the $(n,m)$ code can protect its single encoded qubit from $\lfloor(n-1)/2\rfloor$ $\Z$ errors and $\lfloor(m-1)/2\rfloor$ $\X$ errors, irrespective of gauge. We refer to this procedure as standard correction. 

The $\Z$ gauge codewords can be simply described as follows. 
The $\Z\Z$ gauge operators enforce the parity checks of the ($\Z$-basis) $m$-fold repetition code in each row, so that the codewords must be spanned by $\ket{\ubar{0}}=\ket{0}\otimes \cdots\otimes \ket{0}$ or $\ket{\ubar{1}}=\ket{1}\otimes\cdots\otimes \ket{1}$, where each $\ket{\ubar{i}}$ consists of $m$ qubits.
Now it is easy to build up the codewords recursively by adding rows. 
For $|{\bar i'}\rangle$ a logical $\Z$ codeword of the $(n-1,m)$ code, the codewords of the $(n,m)$ code  are just
$|{\bar 0}\rangle =|{\uz}\rangle\otimes |{\bar 0'}\rangle+\ket{\uo}\otimes |\bar 1'\rangle$ and 
$|{\bar 1}\rangle =\ket{\uz}\otimes |{\bar 1'}\rangle+\ket{\uo}\otimes |\bar 0'\rangle$. 
This can be more succinctly expressed by writing the $(n,m)$ encoded state $\ket{\bar\varphi}$ in terms of the same state $\ket{\bar\varphi'}$ encoded in the $(n-1,m)$ code and its logical $\X$ operator as 
\begin{align}
\label{eq:recursion}
\ket{\bar\varphi}=\ket{\uz}\otimes \ket{\bar\varphi'}+\ket{\uo}\otimes \bar \X'\ket{\bar \varphi'}\,.
\end{align}

\section{Error correction}
\label{sec:ec}
\subsection{Damping of Bacon-Shor codewords}
Duan \emph{et al.}\ observed that the $(t+1,t+1)$ code satisfies the error-correcting conditions to order $t$~\cite{duan_multi-error-correcting_2010}. 
It is important to note that this only holds in $\Z$ gauge. 
Consider the $(2,2)$ code in $\X$ gauge, whose codewords are $\ket{\hat 0}=\ket{0000}+\ket{0101}+\ket{1010}+\ket{1111}$ and $\ket{\hat 1}=\ket{0011}+\ket{0110}+\ket{1001}+\ket{1100}$. 
It is easy to see that the error-correcting conditions are not satisfied for single jumps. 
If the first qubit is damped, the state resulting from $\ket{\hat 1}$ will contain a term $\ket{0001}$, but this term will also be present in the result of damping the second qubit of $\ket{\hat 0}$.
 
Before proceeding to the details of our two correction schemes, let us first consider which error operators are relevant to order $t$. 
Each qubit is afflicted with either Kraus operator $A_0$ or $A_1$, and since damping is linear in $\gamma$ we need only consider products of Kraus operators with at most $t$ factors of $A_1$. 
As $A_0$ effectively contributes phase errors starting at second order in $\gamma$, error operators with $k\leq t$ factors of $A_1$ can be regarded as $k$ damping errors on the corresponding qubits and no more than $\lfloor(t-k)/2\rfloor$ phase flips on the remaining qubits. 
For any given $A_0$ factor we are really only concerned with the first term in the expansion given in \eqref{eq:A0}, since using the higher order terms only reduces the number of possible phase errors on other qubits. 

It is also useful to examine the effect of damping errors on $\Z$-gauge Bacon-Shor codewords. 
Consider the effect of damping on a particular row. 
Since the codewords are constructed from $\ket{\uz}$ and $\ket{\uo}$, by symmetry it suffices to consider damping of the first $k$ qubits by the error operator $E_k=A_1^{\otimes k}\otimes A_0^{\otimes {m-k}}$, for some $0<k\leq m$.
From \eqref{eq:recursion} we have $E_k\ket{\bar\varphi}=\ket{0}^{\otimes k}\otimes A_0^{\otimes {m-k}}\ket{1}^{\otimes m-k}\otimes \bar \X'\ket{\bar\varphi'}$. 
Thus, damping decouples the given row from the codeword and maps its logical information to the Bacon-Shor code with one fewer row, applying a logical bit flip along the way. 
Moreover, it manifests itself by altering at least one of the parity checks in the row. 
Logical $\Z$ is now $-\Z^{\text L}_{(n-1,m)}$, the left column of $\Z$ operators in the $(n-1,m)$ code, and the logical $\X$ is just $\X^{\text T}_{(n-1,m)}$. 
Damping of further rows moves the quantum information into fewer and fewer rows, applying a logical bit flip each time. 

\subsection{Clifford decoder}
\label{sec:decoderA}
In light of the above, the following procedure will recover the quantum information encoded in the $(n,m)$ code to order $\min(n-1,m-1)$ in $\gamma$. 
This is an explicit scheme enabling  the $(t+1,t+1)$ code to be perfectly recovered to order $t$. 
First, the $\Z\Z$ gauge operators are measured to identify rows afflicted with damping errors.
Next the remaining $n'$ undamped rows are treated as an $(n',m)$ codeword, and its $\X$ stabilizers measured. 
After performing standard phase error correction, a logical $\X$ is applied if $n-n'$ is odd. 
The original encoded information can then be obtained by reversing the $(n',m)$ encoding circuit.

To verify the error correcting capability of this scheme, first observe that $E_k$ will produce a state with nontrivial syndrome unless $k=m$, i.e.\ unless all qubits in a row are damped. 
We therefore require $m\geq t+1$ in order to avoid this possibility at order $t$.
Similarly, as the encoded information is lost if at least one qubit in each row is damped, we also require $n\geq t+1$.
Given that $n'$ undamped rows remain, to reverse the action of $A_0$ on each qubit to order $t$, by the usual discretization argument~\cite{shor_scheme_1995,gottesman_introduction_2010}  it is sufficient to correct at most $\lfloor (t-n+n')/2\rfloor$ phase errors. 
Standard phase error correction achieves this for any $n'\leq n$ since we have already chosen $n\geq t+1$. 
Finally, the last step reverses the logical bit flip acquired from each damped row. 

Recovery to order $\min(n-1,m-1)$ is precisely the same performance the code has against erasures~\cite{gottesman_stabilizer_1997}. 
The reason this works for amplitude damping is that we need only detect damping events, not determine precisely which qubits were damped. 
Hence, the bit-flip error-detecting properties of the code suffice. 
Phase errors, meanwhile, are handled using error correction, but there are essentially only half as many phase errors to correct.  
Note that neither this nor any procedure can recover $(t+1,t+1)$ codewords to order $t$ from the twirled channel $\cN_\gamma^{\text{Pauli}}$. 
Doing so would require correcting $t$ bit flips in a single row, which is impossible for the $(t+1)$-bit repetition code. 

\subsection{Clifford codeword correction}
\begin{figure}[th]
\[\Qcircuit @C=1.3em @R=1em {
& \lstick{\ket{\varphi}_A} &  \multigate{2}{\cM_{\Z\Z}} & \gate{\cM_{\X}} &\cctrl{2}\\
&  &\pureghost{\cM_{\Z\Z}} & \cctrl{1}\\
& \lstick{\ket{+}_B}&\ghost{\cM_{\Z\Z}} &\gate{\X}&\gate {\Z}&\rstick{\ket{\varphi}_B} \qw
}
\]
\caption{\label{fig:teleport} One-qubit teleportation circuit.}
\end{figure}
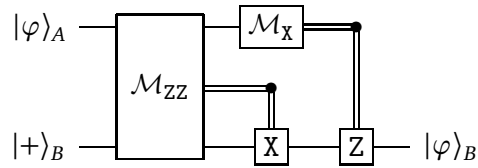

Combining the above decoder with one-bit teleportation~\cite{zhou_methodology_2000} yields a simple Clifford correction circuit which restores the input codeword without completely decoding and re-encoding it. 
This is more amenable to fault-tolerant implementation. 
The action of the one-bit teleportation circuit at the logical level is depicted in Figure~\ref{fig:teleport}, while for encoded states we interchange the order of error correction and teleportation steps as follows. 
%For encoded states, we make use of the freedom to interchange the order of error correction and teleportation steps the procedure for restoring input codewords is as follows.

First prepare $\ket{\bar +}_B$ in the new block $B$.  
Then measure $\tilde \Z$, the product of $\Z$ on qubits in the first column of $A$ and the first column of $B$. 
Next measure the $\Z\Z$ gauge operators on $A$. 
Depending on the result, measure all qubits in undamped rows in the $\X$ basis or the first qubit in damped rows in the $\Z$ basis. % and, depending on the result, individually measure all qubits in damped rows in the $\Z$ basis and all qubits in undamped rows in the $\X$ basis. 
Finally, the necessary Pauli correction operators are determined from the measurement results.
The value of $\bar \Z_A\bar \Z_B$ is given by $\bar \Z_A\bar \Z_B=(-1)^\ell \tilde \Z$, where $\ell$ is the number of rows in which the first qubit is found to be in state $\ket 0$. 
Meanwhile, the value of $\bar \X_A$ is found by multiplying the individual $\X$ measurement results in each undamped row (treating them as $\pm 1$) and then taking the majority. 

To see that this procedure indeed restores the original codeword, first note that the output is certainly a valid codeword, since measurement of $\tilde \Z$ commutes with the $\Z\Z$ gauge operators and $\X$ stabilizers in block $B$. 
Thus, we need only insure that the logical information is properly extracted. 
The logical $\bar \X_A$ operator is determined in the usual way from individual $\X$ measurements and standard error correction. 
Matters are more subtle for logical $\bar \Z_A$. 
Unlike the description above, here it makes sense to express $\bar \Z$ in terms of $\Z^{\text L}_{(n,m)}$ and not in terms of the logical operator of the subcode of undamped rows.  
This is easily done. 
Suppose one row is damped, so that, as above, $\bar \Z=-\Z^{\text L}_{(n-1,m)}$. 
If the first qubit in the damped row is damped, then it is necessarily in the $\Z=+1$ eigenstate $\ket 0$, meaning $\bar \Z=-\Z\otimes \Z^{\text L}_{(n-1,m)}=-\Z^{\text L}_{(n,m)}$. 
On the other hand, if the first qubit was not damped, it is left in the state $\ket 1$ by virtue of being in a damped row, and we have $\bar \Z=\Z^{\text L}_{(n,m)}$. 
A similar argument applies to any number of damped rows, and shows that the procedure above extracts the correct value of $\bar \Z_A\bar \Z_B$ from the measurement of $\tilde Z$.

%undamped qubits in damped rows are left in the state $\ket 1$, so is not damped, then it is 

%The outcome of $\tilde \Z$ differs from $\bar \Z_A\bar \Z_B$ only to the extent that the first qubits of some rows are damped. 
%Since damping of the first qubit in a row will necessarily produce a state $\ket 0$, the logical operator in this case can be expressed as $\bar \Z_A=-\Z^{\text L}_{A}$. 
%Similarly, if undamped, the first qubit Damping of any other qubit in the row leaves the first qubit in the state $\ket 1$, whence $\bar \Z_A=\Z^{\text L}_{A}$. 
%For $\bar \X_A$ we use the standard decoding of the repetition code to account for phase errors from the action of $A_0$ on qubits in undamped rows. 

\subsection{Syndrome correction}
\label{sec:synd}

A simpler, purely syndrome-based correction scheme is also possible, one where the only quantum operations needed are stabilizer measurements and Pauli corrections. 
This method has less noise tolerance, but is capable of correcting $\cN_\gamma^{\text{Pauli}}$ as well as $\cN_\gamma$.  
The basic idea is to use the $\Z\Z$ gauge operators to identify the position of damped qubits, i.e.\ to use the $\Z\Z$ guage operators for error correction and not just detection. 
Correlations between bit and phase errors in $\cN_\gamma^{\text{Pauli}}$ can  be utilized to enable recovery to order $\min(n-1,\lfloor (m-1)/2\rfloor)$, meaning $(t+1,2t+1)$ codewords can be perfectly restored to order $t$.  

The scheme is as follows. 
First, the $\Z\Z$ gauge and $\X$-type stabilizer operators are measured. 
The measurement results of the former are then used to perform standard bit-flip error correction in each damaged row. 
Next, standard phase-flip error correction is performed on the $(n',m)$ code consisting of the undamaged rows; the necessary $\X$-type syndromes are computed from the $\X$ stabilizer measurement results.  
Finally, any suitable pattern of $\I$ and $\Z$ is applied to the first qubits of damaged rows to ensure all remaining $\X$ syndromes are reset to $+1$.

The $\Z\Z$ syndromes enable the decoder to identify and undo the actual bit-flip error pattern as usual, and thererfore recovery to order $t$ requires a code of at least $2t+1$ columns. 
Correction of phase errors, however, does not proceed by attempting to identify the precise phase error pattern.  
Note that in $\Z$ gauge we may as well regard any (odd number of) phase flips in a row as a phase flip on the first qubit, and so phase error correction can be reduced to recovery from the usual one-dimensional repetition code. 
But, and here is the distinction to the usual correction scheme, the error model does not just assume independent noise.
Instead, rows containing bit-flipped qubits are as likely as not to be afflicted by phase flips, while phase flips occur at a lower rate in undamaged rows. 
Hence the damaged rows should simply be excluded in majority-vote decoding.
Therefore, we can recover the quantum codeword by performing error correction on the undamaged rows and then resetting the $\X$ stabilizers involving the remaining rows by action on the damaged rows as needed. 
Following the calculation in \S\ref{sec:decoderA}, this requires at least $t+1$ rows to recover from order $t$ error events.

\section{Fault-tolerant implementation}
\label{sec:ft}

Now we turn to the question of whether the qubit savings afforded by either of the two correction schemes can be realized when damping noise also afflicts the correction operations themselves. 
There are two immediate issues to confront. 
One is the potential for catastrophic error propagation in the correction implementation itself, excluding ancillas. 
The other is how damping noise on ancilla systems and its subsequent propagation affects correctability in future correction cycles.  
We do not attempt a complete analysis of fault-tolerant implementation.
Rather, our goal is to show that neither of these issues is immediately fatal to the proposed correction schemes, so there is indeed hope for realizing the savings offered by adapting the correction to the noise model.

First, though, it is important to note that it is not possible to reach arbitrarily low logical noise rates by choosing large enough $n$ and $m$. 
As with Pauli noise~\cite{napp_optimal_2013}, Bacon-Shor codes have no nontrival noise threshold against amplitude damping even for ideal error correction.
Consider the event in which every row has at least one damped qubit. 
This occurs with probability $(1-(1-\gamma)^m)^n$ for the $(n,m)$ code, since the probability of any qubit in a row being damped is $1-(1-\gamma)^m$. 
Taking $m=n$, we even have $\lim_{n\to \infty} (1-(1-\gamma)^n)^n=1$, which is easily seen by employing l'H\^opital's rule on the logarithm of the probability. 
The increased error correcting power of larger codes is outstripped by the number of likely errors needing correction, and so code concatenation or some combination with another coding scheme is thus required to have a noise threshold. 
Nonetheless, encoding into small codes will result in a lower logical error rate than no encoding at all, just as observed for Pauli noise in \cite{napp_optimal_2013,brooks_fault-tolerant_2013}. 
We leave finding the optimal code size for a given damping rate $\gamma$ as an open question.

\subsection{Catastrophic errors}
Catastrophic errors are not particularly an issue for syndrome correction, as it can be understood in the usual Pauli error framework. 
Indeed, this fact means that we can appeal to standard fault-tolerant methods such as Steane or Knill error correction and passive implementation of Pauli corrections by tracking the ``Pauli frame''~\cite{steane_active_1997,knill_quantum_2005,gottesman_introduction_2010}.
Nonetheless, these methods do not guarantee recovery of a $(t+1,2t+1)$ code to order $t$, as a proper accounting of the contributions from the various possible faults is still needed. 
This issue is investigated in the next section.  

For the Clifford decoder and teleportation correction, however, catastrophic errors are a major concern.
Indeed, damping of even a single qubit of an arbitrarily-sized codeword can cause a logical bit flip error in the Clifford decoder if it occurs between the $\Z\Z$ gauge and $\X$ stabilizer measurements. 
Take the $(2,2)$ code, for instance, and imagine that one of the first two qubits is damped after the $\Z\Z$ operators are found to both have the value $+1$. 
The stabilizer are now $-\Z\Z\I\I$, $\I\I\Z\Z$, and $(-1)^j\Z\I\I\I$, where $j\in\{0,1\}$ indicates which qubit was damped, while the logical operators can be expressed as $\I\I\X\X$ and $-(-1)^j\Z\I\Z\I$. 
Measurement of $\X\X\X\X$ will remove $(-1)^j\Z\I\I\I$ as a stabilizer, leaving us with no access to the value of $j$, and so error-correction fails. 
The difficulty is that the location of damping determines whether the encoded bit value is flipped or not, and the $\X$ stabilizer measurement destroys this information, also in codes of arbitrary size.   
Note that that the syndrome decoder avoids this problem by using the $\Z\Z$ checks to determine where damping occurred. 

A similar problem plagues teleportation correction. 
One source of trouble is damping occurring between then $\tilde \Z$ and $\Z\Z$ measurements. 
We can think of the $\tilde \Z$ measurement as transferring the value of $\Z^{\text L}_A$ to the second block, and this value will be either $\pm \bar \Z$ depending on whether any first qubits are damped before or after the $\tilde \Z$ measurement itself.
But in the scheme we have no way of knowing which is the case. 
This problem can be avoided by measuring $\tilde \Z$ in all the columns and making use of the fact that the logical operator could be defined using any column, not just the first. (This will not unduly disturb $\bar \X_B$ since the product of $\tilde \Z$ in two columns is a $\Z$-type stabilizer.)
After the remaining measurements of the correction cycle, we can determine which column is unaffected by damping and take it as the value of $\bar \Z_A\bar \Z_B$. 
To order $t$ there will definitely be at least one good column, since there are $t+1$ in total. 

Another issue is damping subsequent to the $\Z\Z$ checks. 
This will lead to a random value of the computed $\X$ value of the affected row, and thus behaves as a phase error. 
Hence phase errors now essentially occur at order $\gamma$, not just $\gamma^2$, which implies that a $(2t+1,t+1)$ code will be needed to recover to order $t$. 
Though note that this should be understood as a lower bound on the code size, since we have not considered the effects of multiple damping events distributed over the different stages of the correction procedure. 
We will see in the next section that the additional rows will anyway be needed to combat phase errors arising from damping of ancilla qubits.

\subsection{Error propagation}

The second issue is whether subsequent error correction cycles can deal with the effects of damping noise in the ancillas needed for correction in earlier cycles. 
Let us examine this effect for single-qubit ancilla schemes for measuring the $\Z\Z$ check operators, since this possibility is indeed one appeal of Bacon-Shor codes in the first place. 

The most straightforward measurement setup is to prepare a qubit in the $\ket 0$ state, sequentially apply \textsc{cnot} gates from the two data qubits to the ancilla, and then measure the ancilla in the standard basis. 
However, it is not difficult to work out that damping of the qubit after the first \textsc{cnot} will project the data qubit onto the $\ket 1$ state.
The Kraus operator for this case is simply $(\id\otimes A_1)U_{\textsc{cnot}}(\id\otimes \ket 0)=\sqrt{\gamma}\ketbra 1\otimes \ket 0$, where the first qubit is the control and the second the target. 
This effectively introduces a phase error in the data block, since an otherwise undamaged codeword $\ket{\bar\varphi}$ will be transformed into $\ket{\uo}\otimes \bar \X'\ket{\bar\varphi'}$. 
Logical $\bar \Z$ is unaffected, but both $\X$ stabilizers involving the damaged row will be flipped with probability one half, just as with a phase error. 
Moreover, the output state is such that phase error correction will catch this error and recover.

This same effect occurs in the Pauli noise model, since a damping event is now a bit flip, which is accompanied by a phase error that will propagate into the data block.
And the same conclusion holds for a measurement setup using a single ancilla qubit in the $\ket +$ state and \textsc{cphase} gates. 
%Indeed, this even more directly implies that damping of the ancilla in between the two gates leads to a phase flip on the data block.

Ultimately, in any of these cases, phase errors effectively occur in the data block at order $\gamma$, reducing the error-resilience of the code. 
For the syndrome decoder, this would imply that a $(2t+1,2t+1)$ code is needed, wiping out any advantage over the standard syndrome decoder.
Perhaps this issue can be avoided by using Steane error correction. 
The one-bit teleportation scheme still retains an advantage over the standard decoder, as only a $(2t+1,t+1)$ code is needed.

\section{Conclusions}

We have shown that Bacon-Shor codes offer the possibility of considerably reduced qubit overhead for error correction against amplitude damping noise. 
From a theoretical point of view, it is interesting that the error correcting conditions can be met by a Clifford circuit, since neither is the channel a Pauli channel, nor does the circuit correct its Pauli twirl. 
From a more applied point of view our work invites a more detailed analysis of fault-tolerant implementation to determine if the qubit savings could be realized in a more realistic setting, such as combined amplitude damping and dephasing noise. 
Moreover, as adapting to the amplitude damping channel is relatively simple, this raises the question of whether similar gains can be found for other codes, e.g.\ surface codes. 
In their investigation of optimal decoding of small surface codes~\cite{darmawan_tensor-network_2016}, Darmawan and Poulin recently observed that non-square lattices with effectively half as many qubits performed better than full square lattices. 
It seems plausible that significant overhead reductions are also possible for simple correction algorithms.% are also possible for $\cN_\gamma^{\text{Pauli}}$ as well as $\cN_\gamma$ in this setting. 

\begin{comment}
question of channel-adapted correction can of course be applied to any code,

and our results show that reductions in overhead could be

This is interesting from a theoretical point of view, as the Clifford decoder can even work against non-Pauli channels. B

ut the implications for near-term implementation of quantum information processors is perhaps more interesting. 
- have shown that fault-tolerance implementation is not immediately ruled out. 
- invites a more detailed analysis and simulation. 

In general the question of channel-adapted decoding can be applied to any code. Here we have seen how the structure of amplitude damping interplays with the structure of the code, perhaps useful in other codes such as the surface code. Gains are to be expected even if we target the Pauli twirl for correction. 

%we have shown that even Clifford error correction can help with a non-Pauli channel. 
The teleportation scheme is also interesting for error models which are mixtures of amplitude damping and dephasing--- T1 and T2 errors.

%interesting aspect for simulations of ft gadgets --- the different EC methods might lead to states that are more or less susceptible to damping in the first place. like when an all 1 state is created, versus the case where all zero might have been. in the latter we don't need to worry about damping... 

\end{comment}

\vspace{.75\baselineskip}
\noindent{\bf Acknowledgments.} 
This work was supported by the Swiss National Science Foundation (SNSF) via the National Centre of Competence in Research ``QSIT'', and by the European Commission via the project ``RAQUEL''. 
AP thanks the NWO WISE grant the La Caixa Foundation for support.

\printbibliography[heading=bibintoc,title=References]

\end{document}